\documentclass[twocolumn, superscriptaddress, preprintnumbers,amsmath,amssymb, showpacs]{revtex4}

\allowdisplaybreaks
\allowdisplaybreaks[4]
\usepackage{graphicx,latexsym}
\usepackage{multirow,array,booktabs}
\usepackage{subfigure}
\usepackage[figuresright]{rotating}
\usepackage{color}
\usepackage{CJK}
\usepackage{dcolumn}
\usepackage{bm}
\usepackage[
            pdfstartview=FitH,
            CJKbookmarks=true,
            bookmarksnumbered=true,
            bookmarksopen=true,
            colorlinks=true,
            pdfborder=001,
            linkcolor=blue,
            anchorcolor=blue,
            citecolor=blue
            ]{hyperref}

\begin{document}

\title{Coexistence of planar and aplanar rotations in $^{195}$Tl}

\author{J. Peng}\email{jpeng@bnu.edu.cn}
\affiliation{Department of Physics, Beijing Normal University,
Beijing 100875, China}

\author{Q. B. Chen}\email{qbchen@pku.edu.cn}
\affiliation{Physik-Department, Technische Universit\"{a}t
M\"{u}nchen, D-85747 Garching, Germany}

\begin{abstract}

The chirality suggested for the doublet bands B2 and B2a in
$^{195}$Tl in $A\sim190$ region is reexamined. The potential-energy
curves and the configurations together with the deformation
parameters are obtained by the constrained covariant density
functional theory. The corresponding experimental energy spectra,
energy differences between doublet bands, and the available
$B(M1)/B(E2)$ values are investigated by the fully quantal particle
rotor model. Analysis on the basis of the angular momentum components,
the $K$-plots, and the azimuthal plots suggest a planar rotation
interpretation for the bands B2 and B2a. Hence, it coexists with
the aplanar rotation in the other doublet bands B4 and B4a in
$^{195}$Tl.

\end{abstract}

\pacs{21.10.Re, 21.60.Jz, 21.10.Pc, 27.80.+w}

\maketitle


The phenomenon of spontaneous chiral symmetry breaking is a subject
of general interest. In atomic nuclear physics, it exists in
triaxially deformed nucleus with high-$j$ valence particle(s) and
high-$j$ valence hole(s)~\cite{Frauendorf1997NPA}. Due to the angular
momentum coupling between particle(s), hole(s), and
core, the total angular momentum vector may lie outside the
three principal planes in the intrinsic frame and hence forms as aplanar
rotation. In the laboratory frame, such kind of aplanar rotation
could give rise to a pair of nearly degenerate $\Delta I = 1$ bands
with the same parity, i.e., chiral doublet bands~\cite{Frauendorf1997NPA}.
With the extensive efforts over twenty years, this exotic collective
mode has became a general phenomenon over the nuclear chart. More than
50 chiral candidates were found in odd-odd, odd-$A$, and even-even nuclei
that spread over $A \sim 80$, 100, 130, and 190 mass regions. For more
details, see reviews~\cite{J.Meng2010JPG, J.Meng2014IJMPE, Bark2014IJMPE,
J.Meng2016PS, Raduta2016PPNP, Starosta2017PS, Frauendorf2018PS}
and very recent data tables~\cite{B.W.Xiong2019ADNDT}.

As an important extension of nuclear chirality, the
phenomenon of multiple chiral doublets (M$\chi$D), i.e., having
multiple pairs of chiral doublet bands in a single nucleus, was
predicted and explored extensively by the state-of-art covariant
density functional theory (CDFT)~\cite{Meng2006Phys.Rev.C37303,
Peng2008Phys.Rev.C24309, Yao2009PRC067302, J.Li2011PRC, J.Li2018PRC,
B.Qi2018PRC, J.Peng2018PRC} and observed in
$^{133}$Ce~\cite{Ayangeakaa2013PRL}, $^{103}$Rh~\cite{Kuti2014PRL},
$^{78}$Br~\cite{C.Liu2016PRL}, $^{136}$Nd~\cite{Petrache2018PRC},
$^{195}$Tl~\cite{Roy2018PLB}, and $^{135}$Nd~\cite{B.F.Lv2019PRC}.
These observations confirm the existence of triaxial shapes
coexistence~\cite{Meng2006Phys.Rev.C37303, Ayangeakaa2013PRL,
Petrache2018PRC, Roy2018PLB, B.F.Lv2019PRC}, and reveal the stability
of chiral geometry against the increasing of intrinsic excitation
energy~\cite{Droste2009EPJA, Q.B.Chen2010PRC, Hamamoto2013PRC,
Kuti2014PRL} and octupole correlations~\cite{C.Liu2016PRL}.

As mentioned above, most of observations of chiral doublet bands
concentrate on the medium mass regions. Therefore, there is a
fundamental goal to hunt for new candidates with chirality or
M$\chi$D in new mass regions. On the one hand, the possibility
of chiral doublet bands or M$\chi$D has been investigated in
$^{60}$Ni~\cite{J.Peng2019PLB} and $^{54,56,57,58,59,60}$Co~\cite{J.Peng2018PRC}
in $A\sim 60$ mass region to open the lighter mass area of chirality.
On the other hand, for the heavier mass region, several
high-spin states of chirality have been reported in $A\sim190$
nuclei~\cite{Balabanski2004PRC, Lawrie2008PRC, Masiteng2013PLB,
Masiteng2014EPJA, Masiteng2016EPJA, Roy2018PLB}.

In Ref.~\cite{Roy2018PLB}, the high-spin states in $^{195}$Tl were
reported. Of which, two pairs of doublet bands were assigned to be
built on two different quasi-particle configurations. One of them
is labeled as bands B4 and B4a based on a five-quasiparticle
configuration $\pi i_{13/2}\otimes \nu i^{-3}_{13/2}(pf)^{-1}$, and
the other one bands B2 and B2a based on a three-quasiparticle configuration
$\pi h_{9/2}\otimes \nu i^{-2}_{13/2}$. Based on the systematics of various
experimental observable in $^{191,193,194,195}$Tl isotopes and the triaxial
shapes predicted by the total Routhian surface (TRS) calculations,
it was concluded that both of them are chiral bands. In details, for
bands B4 and B4a, the energies difference between doublets are
small (average separation of 25 keV) and the $B(M1)/B(E2)$ ratios of
the doublet bands are similar. These features do indeed fulfil the
hallmarks of chiral doublet bands~\cite{Frauendorf1997NPA}.

\begin{figure}[!ht]
\includegraphics[width=5.5 cm]{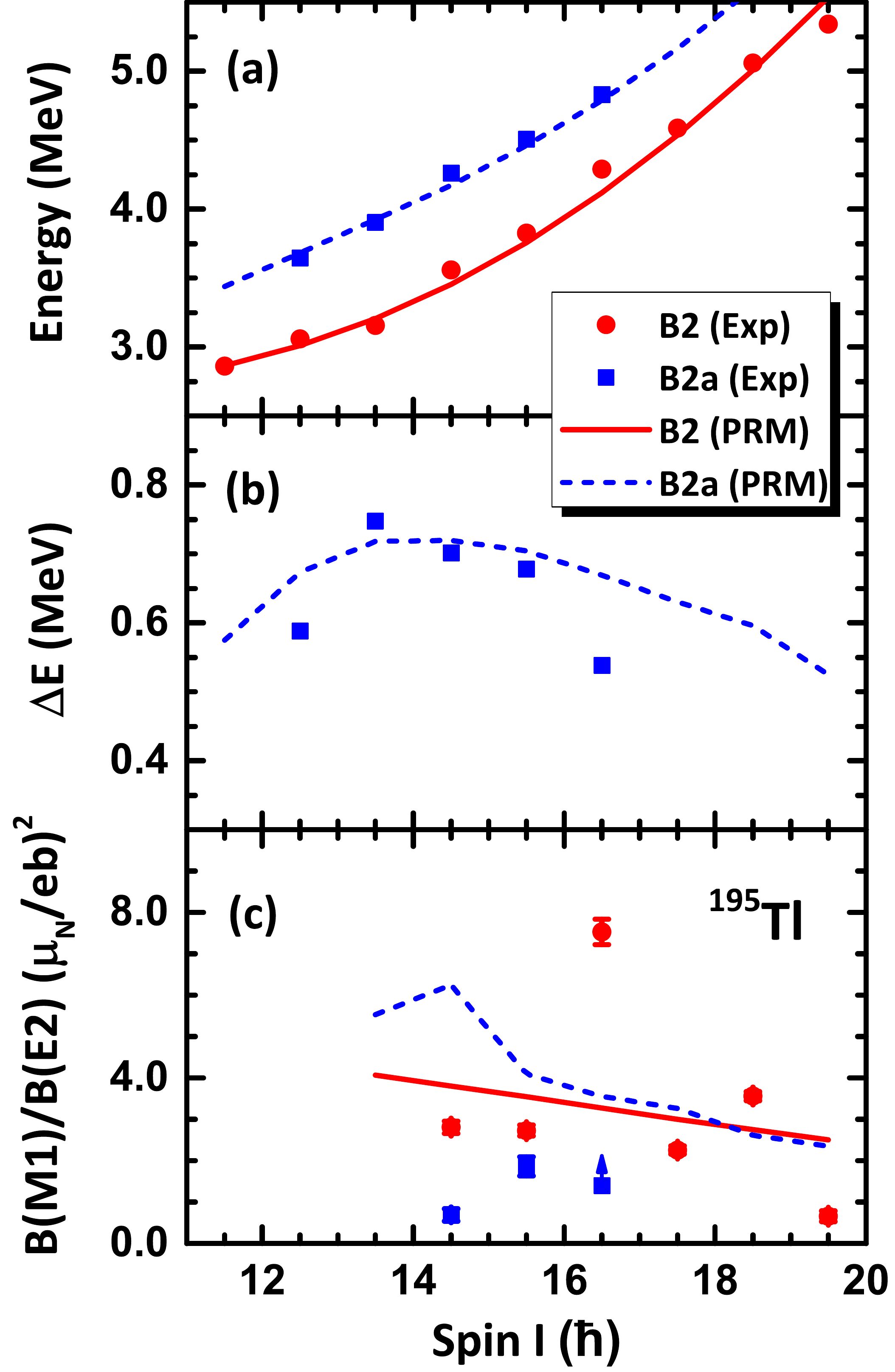}
   \caption{(a) Experimental energy spectra as functions of spin for the bands B2 and B2a in
    $^{195}$Tl in comparison with results calculated by PRM. (b) Experimental and theoretical
    energy differences between the doublet bands. (c) Experimental and theoretical
    $B(M1)/B(E2)$ ratios of bands B2 and B2a.}\label{fig2}
 \end{figure}

For bands B2 and B2a, however, as shown in Fig.~\ref{fig2}, their energy
differences (larger than 500 keV) are much larger than the typical
value 200-300 keV of chiral doublet bands~\cite{B.W.Xiong2019ADNDT} and
their $B(M1)/B(E2)$ ratios are not that similar on the magnitude and
the staggering pattern as a function of spin. This looks as if the interpretation
of chirality for this doublet bands was questionable. Does the aplanar
rotation mode really exist in this doublet bands? To answer this
question, a careful theoretical study of energy spectra, $B(M1)/B(E2)$
ratios, as well as angular momentum behaviors of the core, the proton
$h_{9/2}$ particle, and the neutron $i_{13/2}^{-2}$ holes for bands
B2 and B2a is of necessary.

The aim of the present work is to investigate the chirality in
doublet bands B2 and B2a in $^{195}$Tl in a fully quantal model. As
a quantal model coupling the collective rotation and the
single-particle motions, the particle rotor model (PRM) has been
widely used to describe the chiral doublet bands and achieved major
successes~\cite{Frauendorf1997NPA, J.Peng2003PRC, Koike2004PRL,
S.Q.Zhang2007PRC, B.Qi2009PLB, Lawrie2010PLB,
Shirinda2012EPJA, Q.B.Chen2018PLB, Q.B.Chen2018PRC,
Q.B.Chen2018PRC_v1, Y.Y.Wang2019PLB, Q.B.Chen2019PRC}. In PRM, the total
Hamiltonian is diagonalized with total angular momentum as a good
quantum number, and the energy splitting and quantum tunneling between
the doublet bands can be obtained directly. Furthermore, the basic
inputs for PRM can be obtained from the microscopical constrained
CDFT~\cite{Meng2006Phys.Rev.C37303, J.Meng2016book, Ayangeakaa2013PRL,
Lieder2014PRL, Kuti2014PRL, C.Liu2016PRL, Petrache2016PRC, B.F.Lv2019PRC}.

In this Letter, the constrained CDFT will be first carried out to
obtain the deformation parameters for the assigned valence nucleon
configurations. With these inputs, the PRM will be applied to study
the energy spectra and the electromagnetic transition probabilities
of the doublet bands B2 and B2a in $^{195}$Tl, and to examine their
angular momentum geometries.

The detailed formalism and numerical techniques of the adiabatic and
configuration-fixed constrained CDFT calculation adopted in this
work can be seen in Refs.~\cite{Meng2006Phys.Rev.C37303,
Peng2008Phys.Rev.C24309} and references therein. In the calculations,
the point-coupling density functional PC-PK1~\cite{Zhao2010Phys.Rev.C54319}
with a basis of 12 major oscillator shells is employed, while the pairing
correlations are neglected for simplicity. The constrained calculations
for $\langle \hat{Q}_{20}^2+2\hat{Q}_{22}^2 \rangle \sim
\beta^2$~\cite{Meng2006Phys.Rev.C37303} and quantum number
transformations~\cite{Peng2008Phys.Rev.C24313} are carried out.

\begin{figure*}[!th]
\includegraphics[width=14 cm]{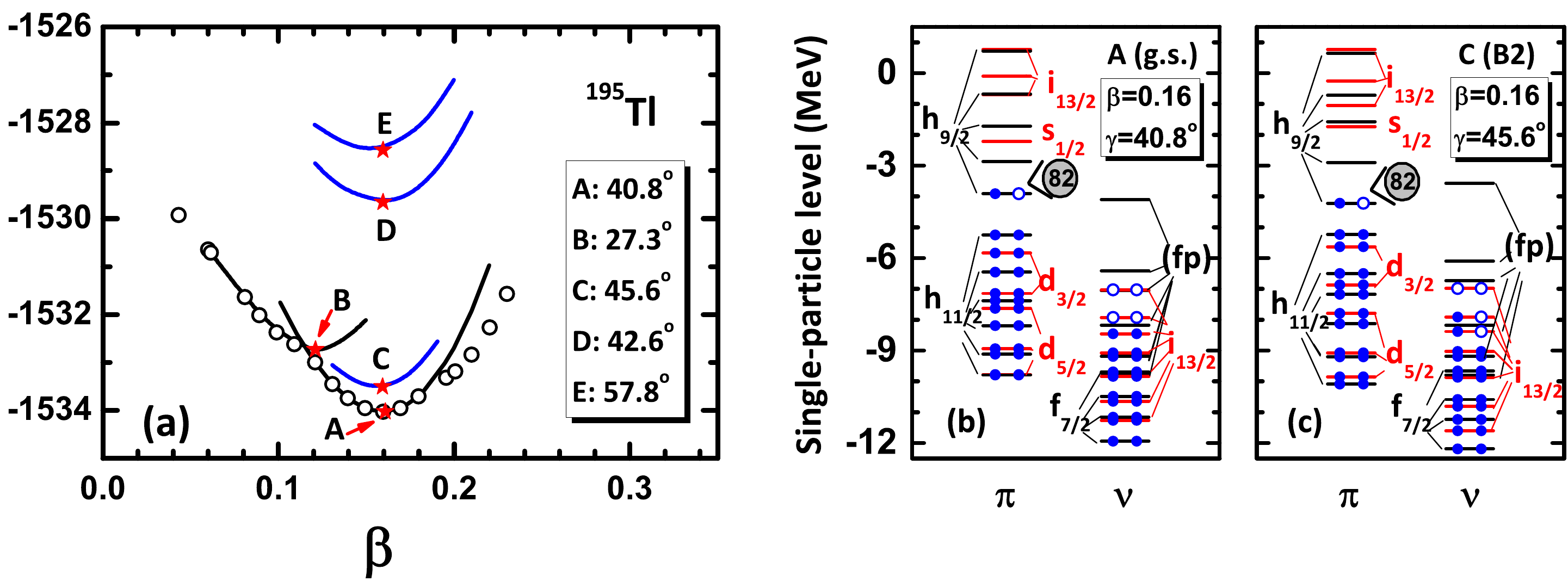}
   \caption{(Color online) (a) The potential-energy curves in adiabatic (open circles)
    and configuration-fixed (solid lines) constrained triaxial CDFT
    calculation with PC-PK1 for $^{195}$Tl. The local minima in the energy
    surfaces for fixed configuration are represented as stars and labeled
    respectively as A, B, C, D, and E. (b) Single-proton (left column) and
    single-neutron (right column) levels near the Fermi surface in $^{195}$Tl for
    the state A. (c) Same as (b) but for the state C.}\label{fig1}
 \end{figure*}

The potential-energy curves for $^{195}$Tl calculated by adiabatic and
configuration-fixed constrained CDFT are presented in Fig.~\ref{fig1}(a),
respectively. In comparison with the irregularities of energy curve
in adiabatic constrained calculations, continuous and smooth energy
curves for each configuration are yielded by the configuration-fixed
constrained calculations. The obvious local minima are marked
by stars and labeled by letters of the alphabet.

Two minima labeled as A and B in potential energy curve in Fig.~\ref{fig1}(a)
have appropriate triaxial deformation, but do not have suitable high-$j$
particle-hole configurations for chirality. Here, state A represents the
ground state, with the deformation parameters $(\beta=0.16,~\gamma=40.8^\circ)$
and the valence nucleon configuration, as shown in Fig.~\ref{fig1}(b),
$\pi h_{9/2}\otimes \nu [i^{-4}_{13/2}(pf)^4]$. Although state
B has the remarkable triaxial deformation, it has only a unpaired nucleon
configuration $\pi 3s_{1/2}$ instead of necessary high-$j$ particle-hole
configuration. By keeping always two aligned neutrons in the fifth and sixth
levels of the $i_{13/2}$ shell, and the other neutrons filling in the orbitals
according to their energies, low-lying particle-hole excitation
state C (the unpaired nucleon configuration $\pi h_{9/2}\otimes \nu
i^{-2}_{13/2}$) are obtained as shown in Fig.~\ref{fig1}(c). By exciting
one proton occupying the $h_{9/2}$ orbital to the $i_{13/2}$ orbital and one
neutron from the $(pf)$ shell to the $i_{13/2}$ shell on the basis of ground
state A, the configuration of state D (the unpaired nucleon configuration
$\pi i_{13/2}\otimes \nu [i^{-1}_{13/2}(pf)^1]$) is obtained.
Similarly, the unpaired nucleon configuration $\pi i_{13/2}\otimes
\nu [i^{-3}_{13/2}(pf)^1]$, labeled E in Fig.~\ref{fig1}(a), are
obtained by exciting one $i_{13/2}$ neutron ($m_z\sim 9/2$) to
a higher $i_{13/2}$ orbit ($m_z\sim13/2$) on the basis of
state D. In the subsequent calculations with the fixed
configurations of C, D and E, the occupations of the valence
nucleons are traced by the configuration-fixed constrained
calculations~\cite{Meng2006Phys.Rev.C37303}. The obtained results
are presented in Figs.~\ref{fig1}(a). Both minima C and D
have deformation parameters $\beta$ and $\gamma$
suitable for chirality, which are C ($\beta=0.16$,
$\gamma=45.6^\circ$) and D ($\beta=0.16$, $\gamma=42.6^\circ$).
These two states have triaxial deformations as well as high-$j$
particle-hole configurations that suitable for establishing chiral
rotation. Therefore, the existence of chiral
doublets or M$\chi$D could be expected in $^{195}$Tl.

It should be noted that the configuration $\pi h_{9/2}\otimes \nu
i^{-2}_{13/2}$ suggested for the doublet bands B2 and B2a in
$^{198}$Tl in Ref.~\cite{Roy2018PLB} corresponds to that of state C
here shown in Fig.~\ref{fig1}(a).
The obtained deformation parameters are close to the one ($\beta=0.15$, $\gamma=39.0^\circ$) by TRS
calculations~\cite{Roy2018PLB}.
To display more clearly, in
Figs.~\ref{fig1}(b) and (c), we show the single-particle energy levels
of proton and neutron near the Fermi surface for the ground state
A and state C, respectively. For the ground state A, we solve the Dirac
equation by filling, in each step of the iteration, the proton and
neutron levels according to their energies from the bottom of the
well. As shown in Fig.~\ref{fig1}(b), for the proton single-particle
energy levels, there is always a particle sitting on the bottom of
the $h_{9/2}$ shell. In comparison, there are
four paired neutron holes sitting on the top of the $i_{13/2}$
shell. For the ground state, the proton has already played a role of
high-$j$ particle, but there is not high-$j$ hole involved. For state C,
two of the neutron holes align along the $z$-axis through one-particle-one-hole
neutron excitation from the ($i_{13/2},~m_z\sim 9/2$) orbital to the
higher ($i_{13/2},~m_z\sim 11/2$) orbital. This kind of excitation
leads to the valence nucleon configuration with the form of $\pi
h_{9/2} \otimes \nu i^{-2}_{13/2}$ shown in Fig.~\ref{fig1}(c).

Subsequently, the PRM calculations with the configuration $\pi h_{9/2}\otimes \nu
i^{-2}_{13/2}$ for the doublet bands B2 and B2a in $^{198}$Tl are performed.
As discussed above, the deformation parameters $\beta=0.16$ and $\gamma=45.6^\circ$
for this configuration at the bandhead were obtained from the
configuration-fixed constrained CDFT calculations. The irroational flow type of
moment of inertia $\mathcal{J}_k=\mathcal{J}_0\sin^2(\gamma-2k\pi/3)$ with
$\mathcal{J}_0 =20.0~\hbar^2/\textrm{MeV}$ and Coriolis attenuation factor
$\xi=0.95$ are adopted according to the experimental energy spectra.
For the electromagnetic transitions, the empirical intrinsic quadrupole moment
$Q_0=(3/\sqrt{5\pi})R_0^2 Z\beta$, and gyromagnetic ratios for rotor
$g_R=Z/A$ and for nucleons $g_{\pi(\nu)}=g_l+(g_s-g_l)/(2l+1)$
($g_l=1(0)$ for protons (neutrons) and
$g_s=0.6g_{s}(\textrm{free})$)~\cite{Ring1980book}
are used.

The calculated energy spectra for the doublet bands B2 and B2a in
$^{195}$Tl are presented in Fig.~\ref{fig2}(a), together with the
corresponding data. The experimental energy spectra are reproduced
excellently by the PRM calculations. Being a quantum model, PRM is
able to reproduce the energy splitting for the whole observed spin
region. It is seen in Fig.~\ref{fig2}(b) that the trend for the
$\Delta E$ between partner bands is reproduced well. Namely, the
$\Delta E$ increases firstly before $I=14.5\hbar$, and then
decreases. Furthermore, as mentioned above, the $\Delta E$ is considerably
large. In detail, the minimum of $\Delta E$ is 588 keV, and the maximum
is 747 keV; much larger than the typical value 200-300 keV of chiral
doublet bands~\cite{B.W.Xiong2019ADNDT}.

Besides the close excitation energies, another important characteristics
of chirality is the similar behavior of the $B(M1)/B(E2)$ ratios for
chiral pairs~\cite{Frauendorf1997NPA, B.W.Xiong2019ADNDT}. In Fig.~\ref{fig2}(c),
one has already observed the significant differences between the experimental
$B(M1)/B(E2)$ of doublet bands, including the magnitudes and odd-even staggering
pattern. As a comparison, the calculated $B(M1)/B(E2)$ values by PRM
are also shown in Fig.~\ref{fig2}(c). No obvious odd-even staggering of the
$B(M1)/B(E2)$ values is presented, although an effect is apparent experimentally.
Except for $I=16.5\hbar$, the PRM calculations show a good agreement
with the data for the band B2. For the band B2a, the PRM overestimates the experimental
data. As discussed in Ref.~\cite{Lawrie2008PRC} for the doublet bands in $^{198}$Tl,
a so-called residual proton-neutron interaction has an effect on the staggering
behavior of $B(M1)/B(E2)$. Here, the discrepancies between the theoretical and
experimental $B(M1)/B(E2)$ might be attributed to the fact that we here do
not consider the proton-neutron interaction.

\begin{figure}[!ht]
  \begin{center}
    \includegraphics[width=8.5 cm]{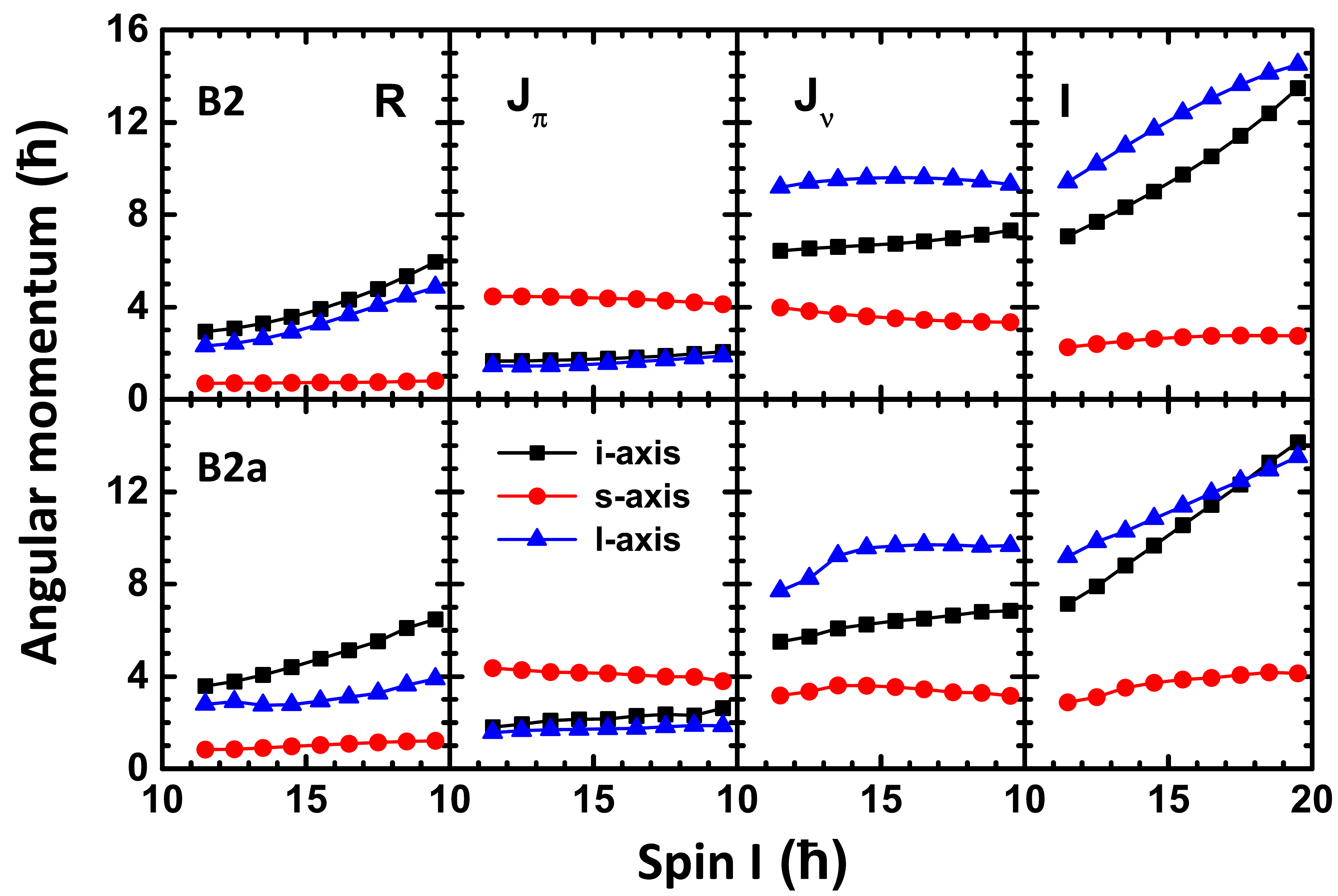}
    \caption{The root mean square components along the short ($s$-, squares),
    intermediate ($i$-, circles), and long ($l$-, triangles) axes for the rotor,
    valence proton, valence neutrons, and total angular momenta as
    functions of spin calculated by PRM for the doublet bands B2
    and B2a in $^{195}$Tl.}\label{fig3}
  \end{center}
\end{figure}

The consistency with the energy spectra of the doublet bands, together
with the discrepancy between the theoretical $B(M1)/B(E2)$ and the data,
motivate us to examine the angular momentum geometry microscopically.
From the eigenfunctions calculated by PRM, one can calculate the
expectation values of the squared angular momentum components along
the short ($s$-), intermediate ($i$-), and long ($l$-) axes for
the rotor, valence proton, valence neutrons, and total angular
momenta, which are shown in Fig.~\ref{fig3}.

The asymmetric degree of the configuration $\pi h_{9/2}\otimes \nu
i^{-2}_{13/2}$ will lead to considerable difference between angular
momenta coming from the valence particle (along the $s$-axes) and
holes (along the $l$-axes), and hence drives triaxial deformation to
deviate away from maximal triaxiality $\gamma=30^\circ$. This will
in return influence the angular momentum geometry of the rotating
system.

As shown in Fig.~\ref{fig3}, for both bands B2 and B2a, the
valence proton in the $h_{9/2}$ orbital contributes about 4.5$\hbar$
to the angular momentum along the $s$-axis, while the valence neutron
$i_{13/2}$ holes contribute to the angular momentum with $\sim 10\hbar$
mainly aligning along $l$-axis as a comparison. The neutron angular momentum
comes mainly from the two aligned neutron holes in the middle of $i_{13/2}$
shell with $m_z \sim -11/2$ and $m_z \sim -9/2$. As a
result, besides the substantial components in $l$-axis ( $\sim 10\hbar$),
the components along $i$-axis ($\sim 6\hbar$) and $s$-axis ( $\sim 4\hbar$)
are also non-negligible.

The angular momentum of collective core has largest component
along the $i$-axis in the whole spin region, because it has
the largest moment of inertia. It should be noted that for both bands,
the rotor angular momenta exhibit also substantial contributions to
the $l$-component. This is understood as reasons of triaxial
deformation deviating far away from $\gamma=30^\circ$ and the strong
Coriolis effects. In details, for the adopted triaxiality $\gamma=45.6^\circ$,
the ratios among the moments of inertia for the three principal axes are
$\mathcal{J}_i:\mathcal{J}_s:\mathcal{J}_l$ =1.00~:~0.07~:~0.54.
Therefore, the collective angular momentum vector no longer tends
to align purely along the intermediate axis, and this leads to the
substantial contributions to the $l$-component. Moreover, the significant
$l$-component of the neutron holes reproduces a strong Coriolis force
to the rotor and drive it align along the $l$-axis to minimize the
energy.

As a result of the angular momentum couplings among the rotor,
valence proton, and valence neutron holes, the total angular momentum
has large components along the $l$- and $i$- axes, while very small
one along the $s$-axis. Such orientations form the angular momentum
geometry of planar rotation in the $l$-$i$ plane instead of chiral
geometry of aplanar rotation.

\begin{figure*}
\includegraphics[width=5.0 cm]{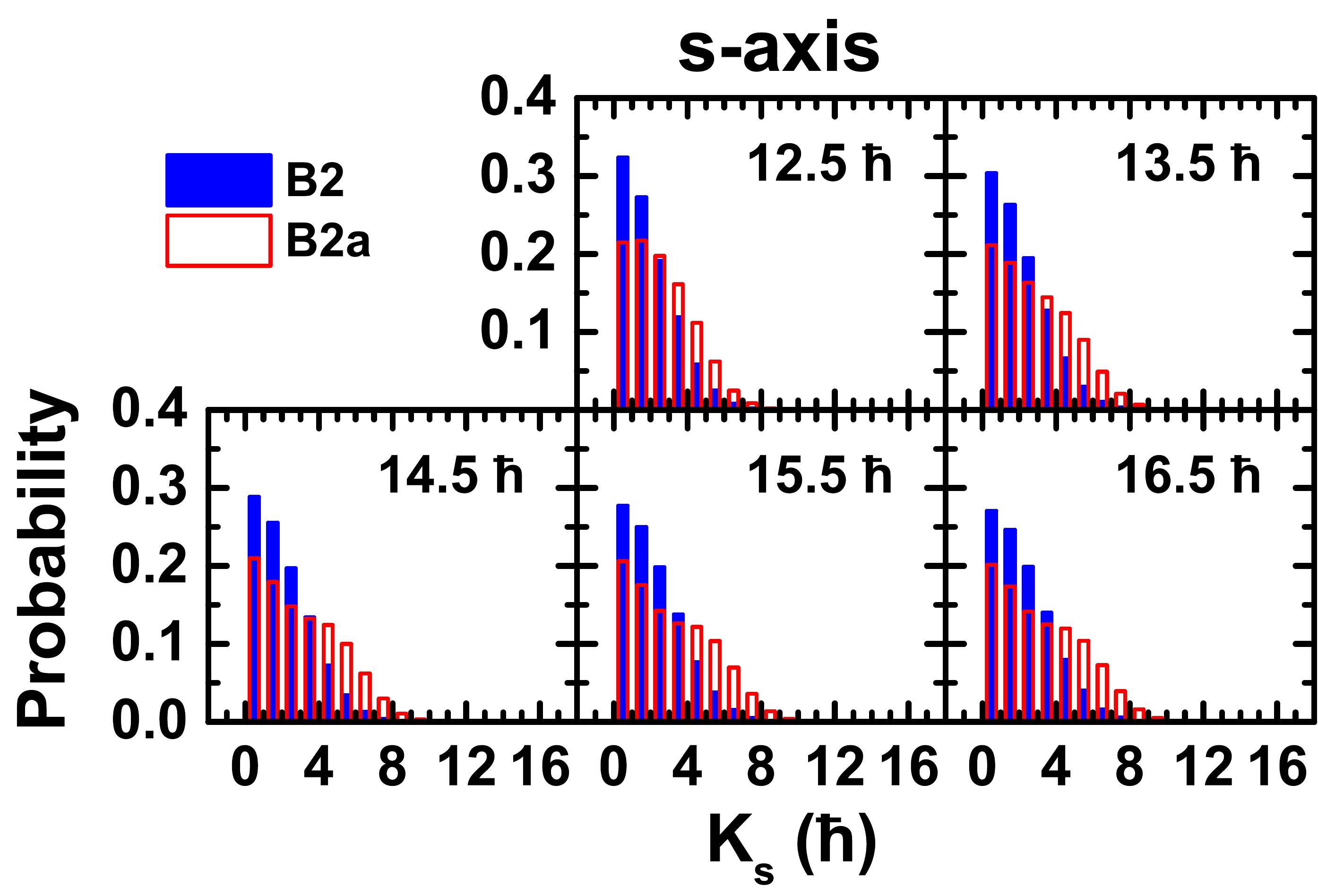}~~
\includegraphics[width=5.0 cm]{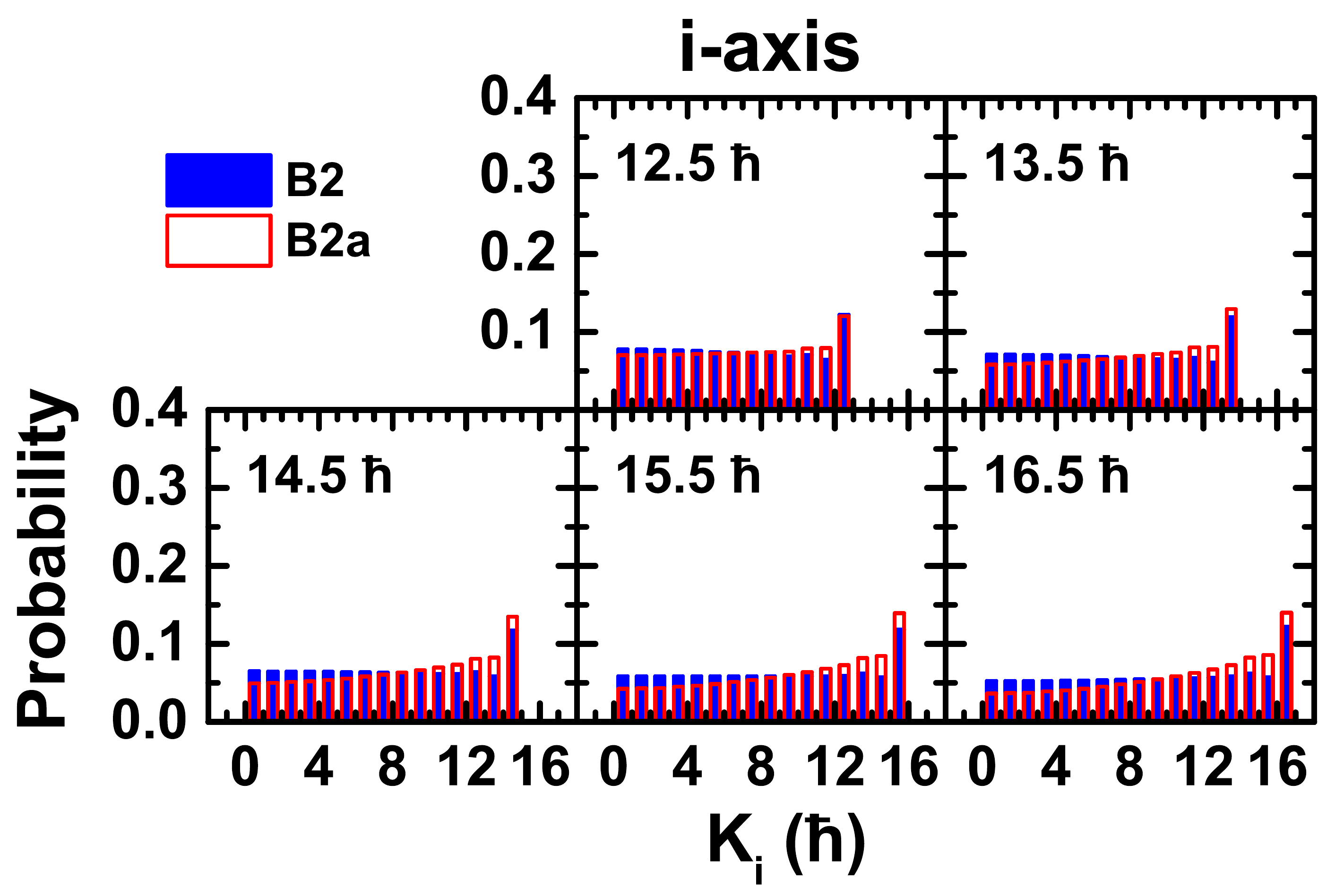}~~
\includegraphics[width=5.0 cm]{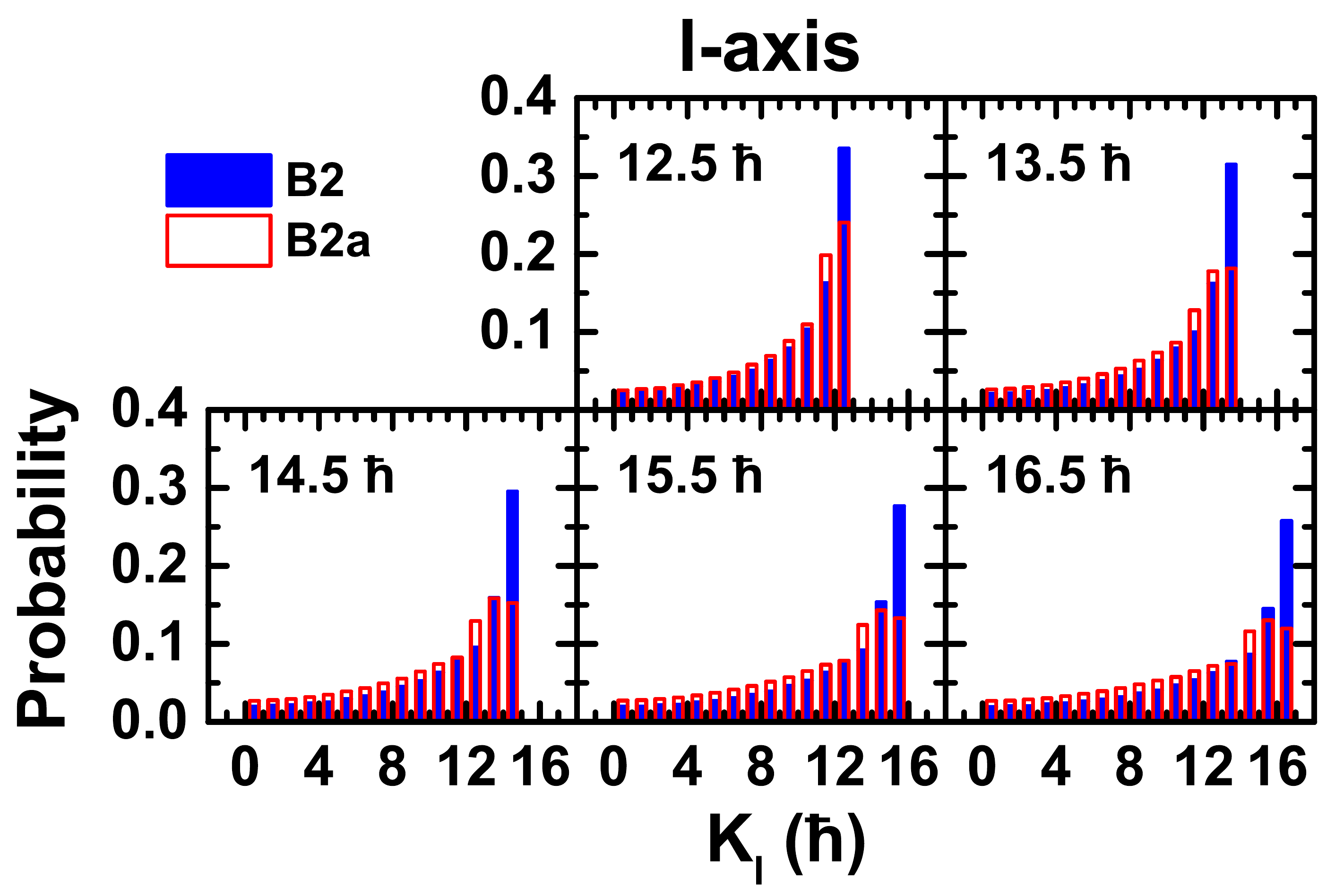}\\
~~\\
\includegraphics[width=5.0 cm]{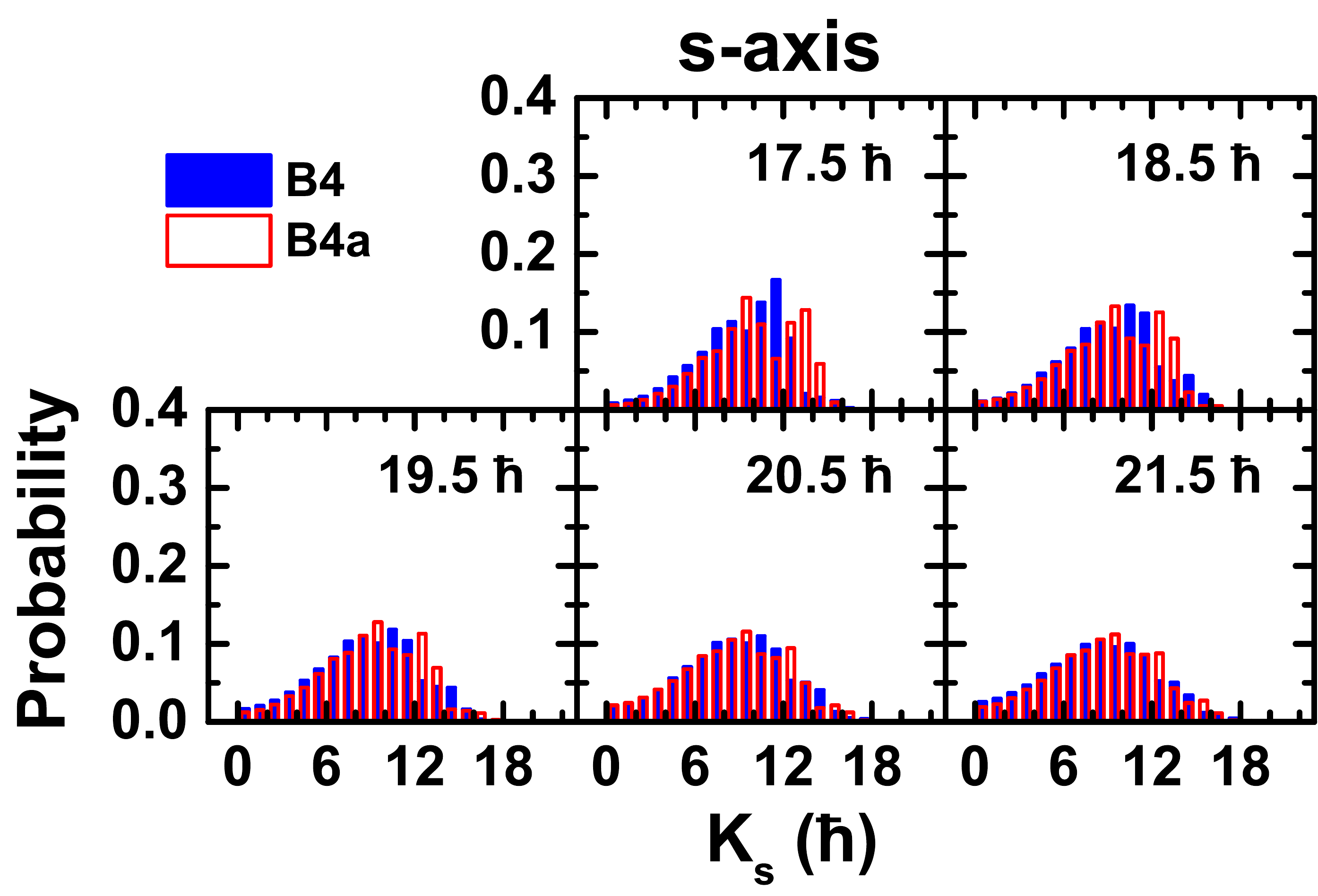}~~
\includegraphics[width=5.0 cm]{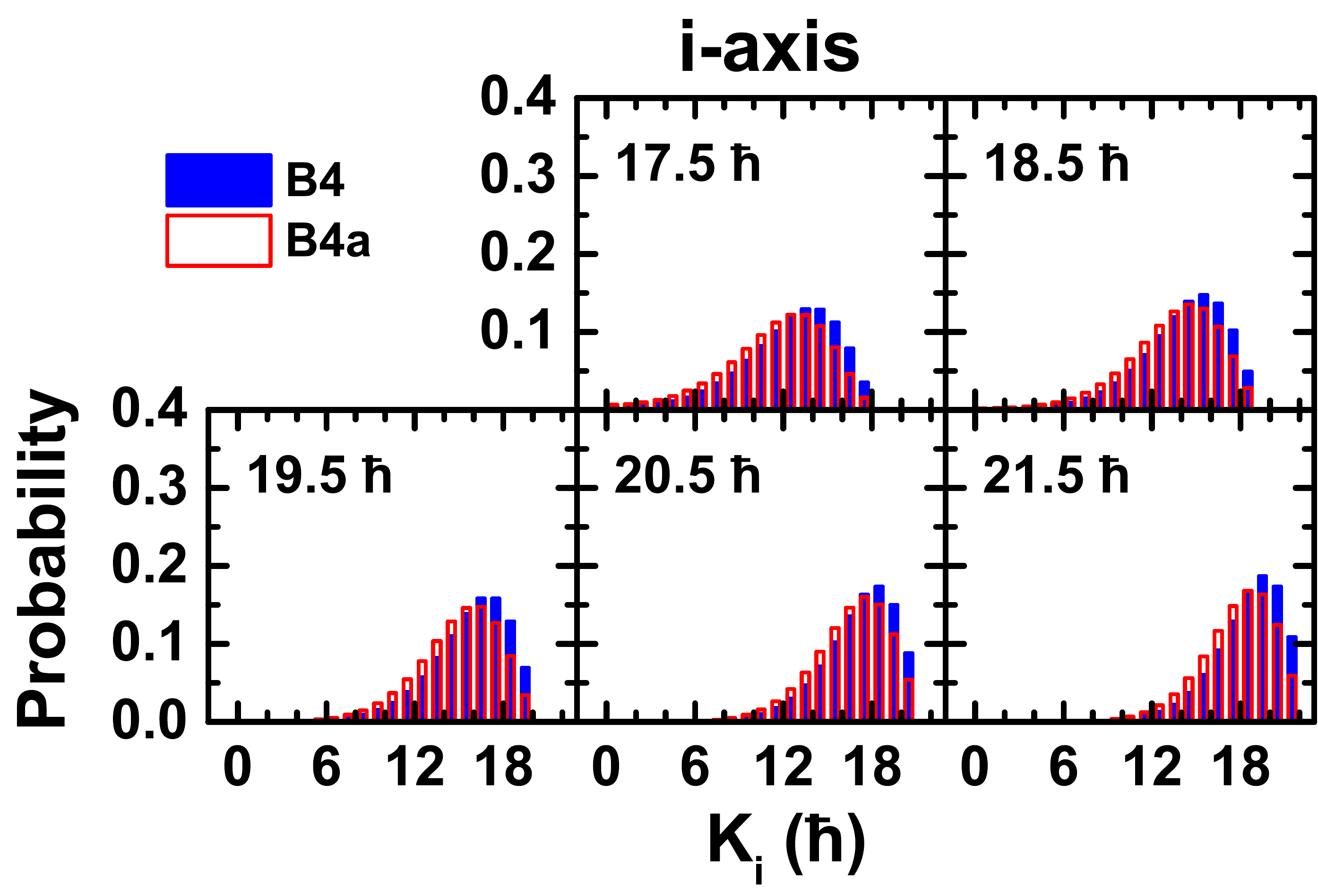}~~
\includegraphics[width=5.0 cm]{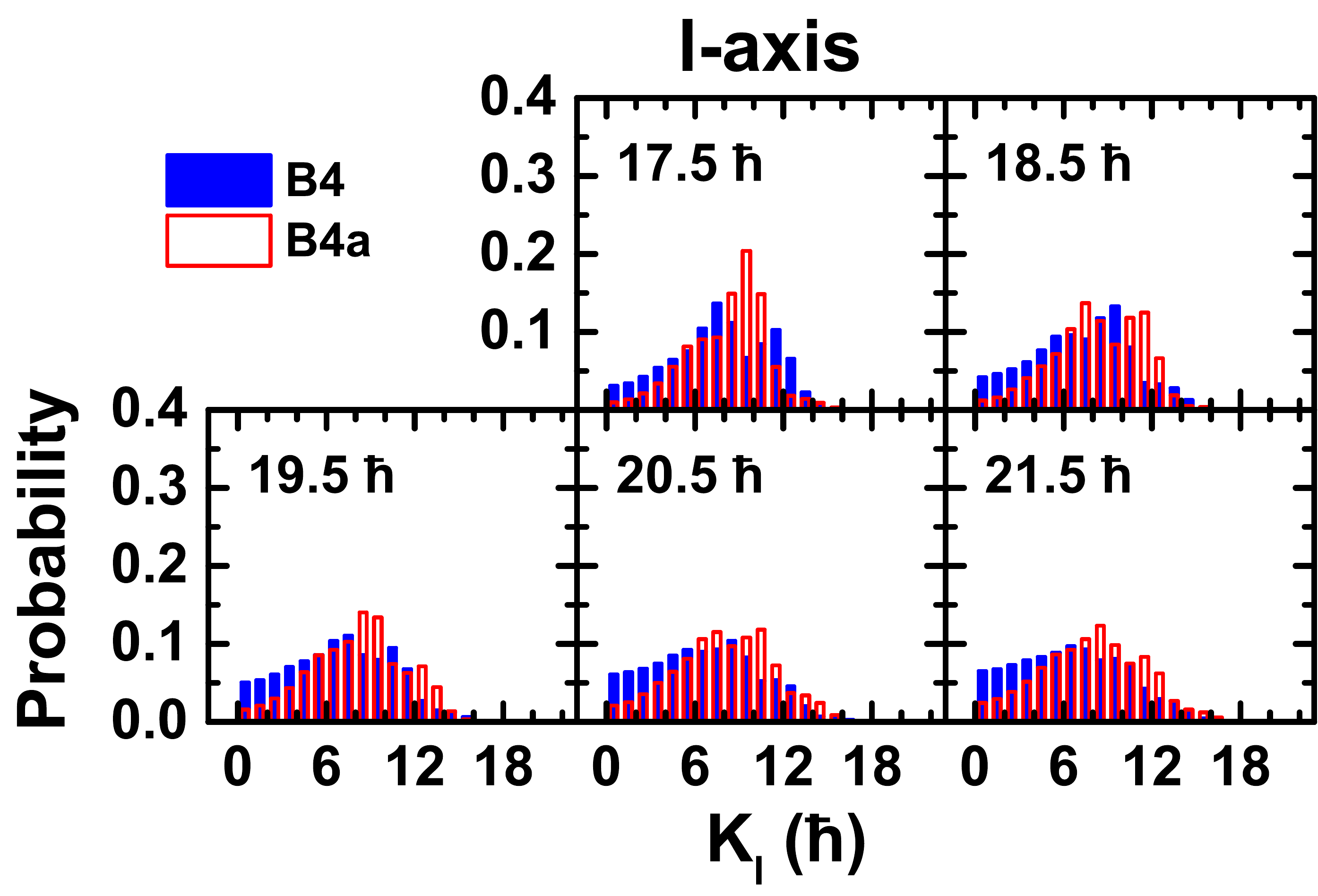}
   \caption{The $K$-plots, i.e., $K$-distributions for the angular momentum
    on the short ($s$-), intermediate ($i$), and long ($l$-) axes calculated by PRM
    for the doublet bands B2 and B2a as well as B4 and B4a in $^{195}$Tl.}\label{fig4}
 \end{figure*}

To further understand the evolution of the angular momentum geometry
with spin, in Fig.~\ref{fig4}, the $K$-plots, i.e.,
$K$-distributions for the angular momentum on the $s$-, $i$-, and
$l$- axes calculated by PRM for the doublet bands B2 and B2a in
$^{195}$Tl are displayed. As seen in the figures, for the whole
observed spin region, the $K$-distributions of bands B2 and B2a
remain almost unchanged.

For bands B2 and B2a, the peaks of $K_s$ locate at $K_s=0.5\hbar$ for
the whole spin region. The peaks locating at the smallest $K_s$ value
indicate that the $s$-components of the proton particle and neutron holes
are canceled each other out. Due to this canceling, the total angular
momentum has too small component along the $s$-axis to form the
aplanar rotation.

The $K_i$-distributions of bands B2 and B2a are similar and spread widely.
Due to the contribution from the neutron holes, there is a weak peak
on the top of dispersive $K_i$-distributions which moves from
$K_i = 12.5\hbar$ to $K_i = 16.5\hbar$ gradually with spin.

The $K_l$-distributions of bands B2 and B2a behave in a similar way,
except that the value of peak of $K_l$ distribution for band B2 is a
bit larger than those for band B2a. Their peaks locate around the
largest $K_l$ value, mainly contributing from the two high-$j$ neutron
holes.

Therefore, the angular momenta for bands B2 and B2a always stay within the
$i$-$l$ plane. The appearance of static chirality or chiral vibration
for bands B2 and B2a are not supported.

\begin{figure}[!ht]
  \begin{center}
    \includegraphics[width=8.6 cm]{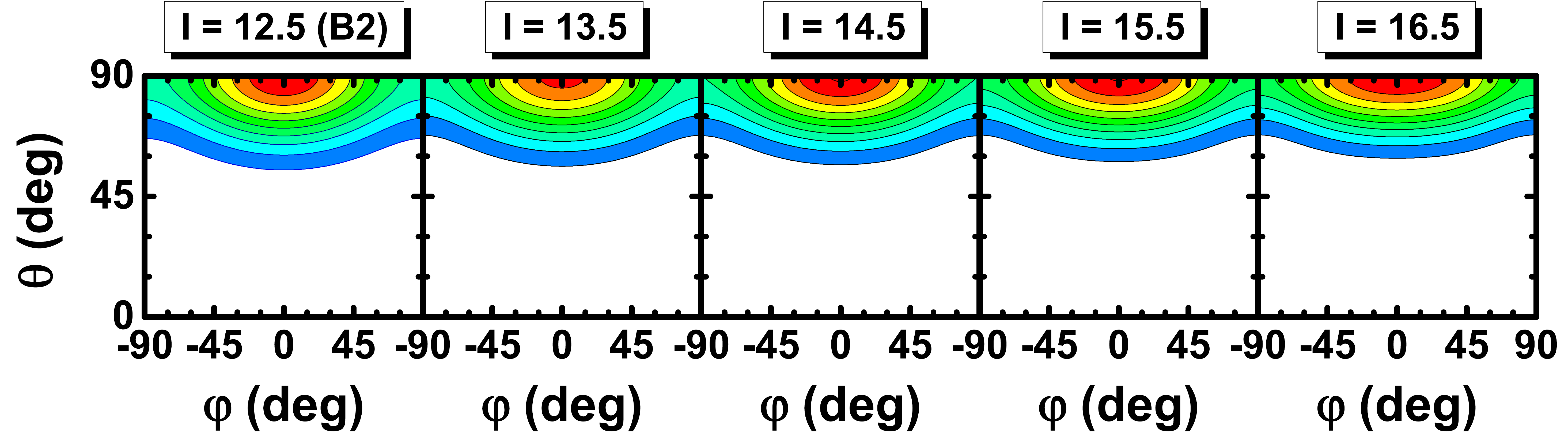}\\
    ~~\\
    \includegraphics[width=8.6 cm]{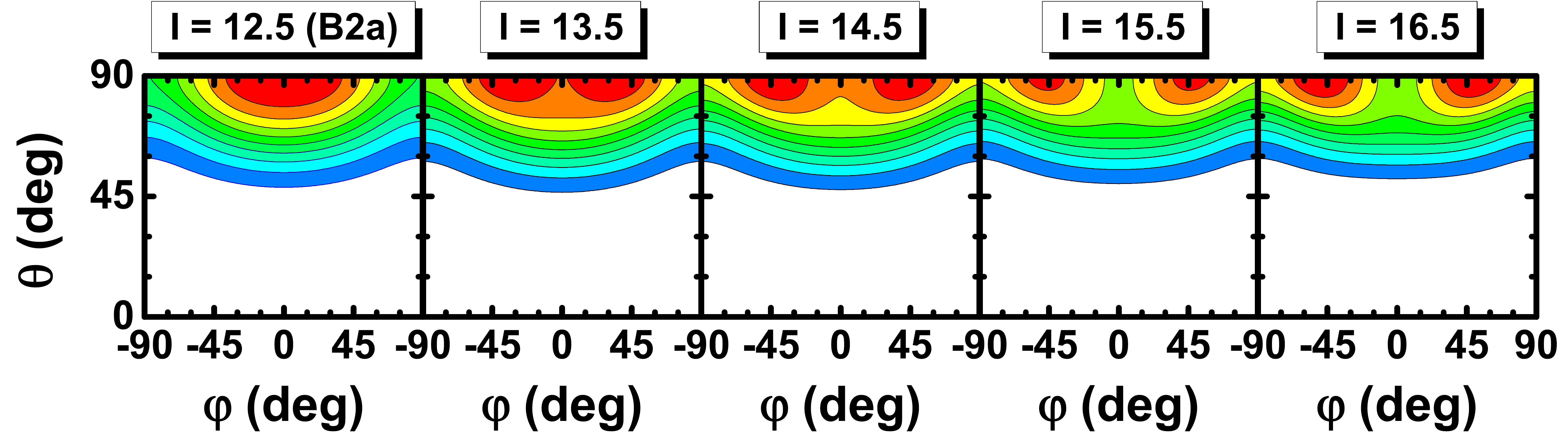}\\
    ~~\\
    \includegraphics[width=8.6 cm]{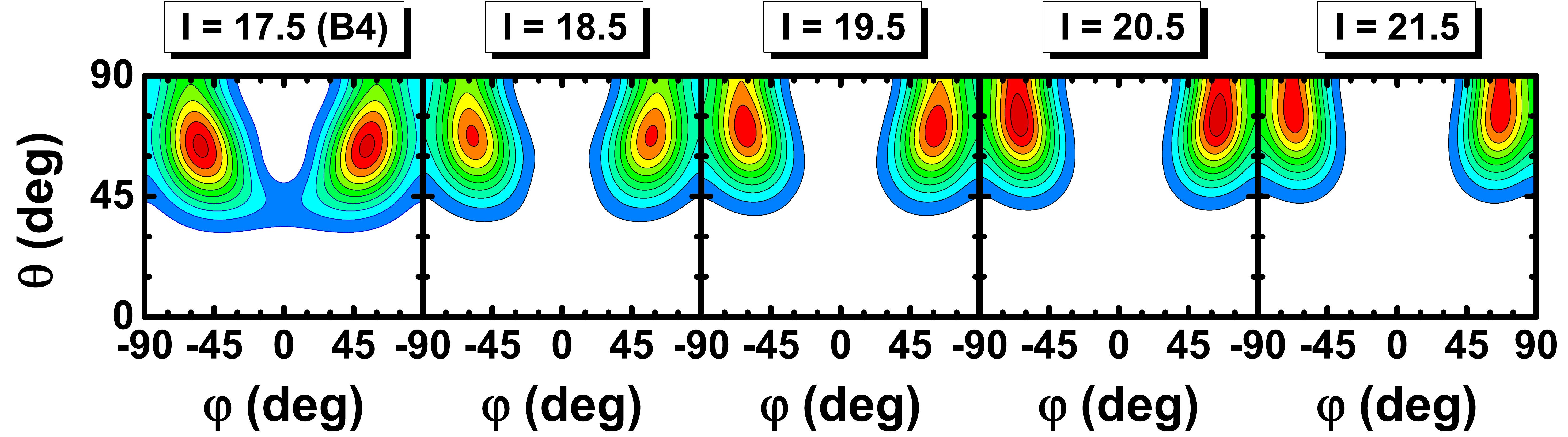}\\
    ~~\\
    \includegraphics[width=8.6 cm]{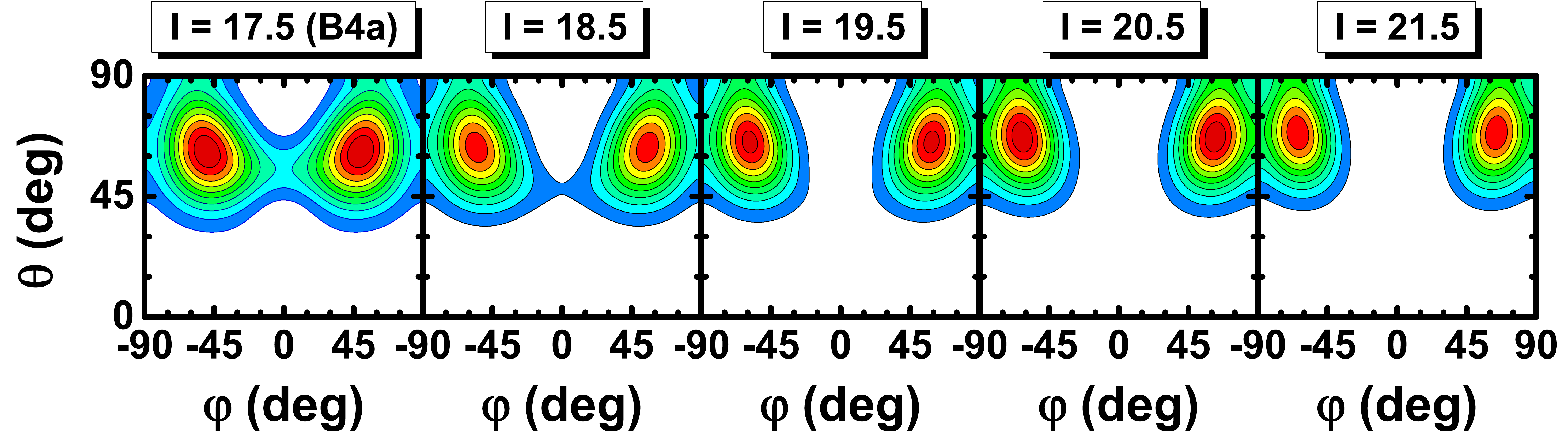}\\
    \caption{The azimuthal plots, i.e., probability density profiles for the orientation
    of the angular momentum on the $(\theta, \varphi)$ plane calculated by PRM for the
    doublet bands B2 and B2a as well as B4 and B4a in $^{195}$Tl.}\label{fig5}
  \end{center}
\end{figure}

In order to visualize the angular momentum geometry in the intrinsic frame,
the azimuthal plots~\cite{F.Q.Chen20017PRC, Chen2018Phys.Rev.C31303,
Streck2018PRC, J.Peng2019PLB, Q.B.Chen2019PRC_v1}, i.e., probability density profiles
$\mathcal{P}(\theta,\varphi)$ for the orientation of the angular
momentum on the $(\theta, \varphi)$ plane calculated by PRM are shown
in Fig.~\ref{fig5} for the doublet bands B2 and B2a in $^{195}$Tl at
$I=12.5$-$16.5\hbar$. Here, $\theta$ is the angle between
the total spin $\bm{I}$ and the $s$-axis, and $\varphi$ is
the angle between the projection of $\bm{I}$ onto the $li$-plane and
the $l$-axis. As shown in Fig.~\ref{fig5}, the azimuthal plots
are symmetric with respect to $\varphi=0^\circ$ due to the D$_2$
symmetry of triaxial shape. The the maximum of the profiles is always
located at $\theta=90^\circ$. This is consistent with the very small
$s$-component of the total spin as shown in Fig.~\ref{fig3}, and rules
out the possibility of the aplanar rotation interpretation.

For the whole observed spin region, the profiles for the orientation
of the angular momentum for band B2 behave in a similar way. The most probable
probability appears at $(\theta=90^\circ, \varphi=0^\circ)$. The profiles also
show soft along the $\varphi$-direction, which is due to the competition from
the $i-$ component of the angular momentum from the rotor and neutron holes.
This feature corresponds to a planar rotation within the $i$-$l$ plane.
Such orientation does not form a chiral geometry and no chirality is
shown in band B2. This is consistent with the angular momentum components
shown in Fig.~\ref{fig3} and $K$-distributions shown in Fig.~\ref{fig4}.

For band B2a, its profile exhibits the similar features as band B2 one
at $I=12.5\hbar$, i.e., a peak at $(\theta=90^\circ, \varphi=0^\circ)$
and softness with respect to the $\varphi$-direction. With the increase
of spin, the higher excitation energies and the stronger Coriolis force
from the valence neutron holes and rotor drive the angular momentum to
align gradually toward the $i$-axis. Two peaks appear, which are
$(\theta = 90^\circ, \varphi \sim \pm 35^\circ)$ for $I=13.5\hbar$,
and $\varphi$ increase gradually with spin. This further supports that
the total angular momentum stays within the $i$-$l$ plane. These features
are quite similar as those for the transverse wobbling shown in
Refs.~\cite{Streck2018PRC, Q.B.Chen2019PRC_v1}, only there the probability
density at $\varphi=0^\circ$ vanishes due to the anti-symmetric $n=1$ wobbling
phonon excitation. The non-vanished probability density at $\varphi=0^\circ$
here is attributed to the significant $i$-component of the neutron holes angular
momenta.

It should be mentioned that the same conclusion is obtained by PRM
calculation based on ($\beta=0.15$, $\gamma=39^\circ$), which is
predicted by TRS of band B2 in Ref.~\cite{Roy2018PLB}.

In addition, the same calculations were also made for the doublet
bands B4 and B4a, and the results of $K$-plots and azimuthal-plots
are also shown in Fig.~\ref{fig4} and \ref{fig5}, respectively
(Noted that in Fig.~\ref{fig5} for bands B4 and B4a, the $\theta$ is
defined as the angle between the total spin $\bm{I}$ and the $l$-axis,
and the $\varphi$ is defined as the angle between the projection of
total spin $\bm{I}$ onto the $s$-$m$ plane and the $s$-axis).
It is clearly seen that these results support aplanar
coupling of angular momenta for this pair of doublet bands.
Namely, the $K$-distributions of the doublet bands B4 and B4a are
similar. Furthermore, the peaks of the three $K$-distributions locate all at
non-zero $K$-values, indicating that the total angular momentum has
finite components along the three principal axes. Two peaks corresponding
to aplanar orientations of the total angular momentum are found in the
azimuthal plots for both of doublet bands at each spin. All of these features
could be understood as a realization of static chirality, and hence give
the small energy differences as the experimental shows. Therefore, the
aplanar (bands B4 and B4a) and planar (bands B2 and B2a) rotations coexists
in triaxial deformed nucleus $^{195}$Tl.

In summary, the chirality suggested in doublet bands B2 and B2a in
the odd-$A$ $^{195}$Tl in $A\sim190$ region is reexamined by
adopting the microscopic constrained CDFT and fully quantal PRM.
The potential-energy curves, the configurations for doublet bands
B2 and B2a of interest together with the corresponding deformation
parameters are obtained by the adiabatic and configuration-fixed constrained
CDFT calculations. The experimental energy spectra and energy differences
between the doublet bands are reproduced well. The available $B(M1)/B(E2)$
values for the bands B2 and B2a are of different odd-even staggering pattern
and exhibit different behaviors. Except for $I=16.5\hbar$, the
$B(M1)/B(E2)$ values for band B2 are reproduced well by the PRM
calculations. The analysis based on the angular momentum components,
the $K$-plots, and the azimuthal plots suggest that a planar rotation
within the $i$-$l$ plane for the bands B2 and B2a. The $s$-component of
angular momentum contributed by the proton particle are canceled out
by the neutron holes. The wobbling like mode for band B2a illuminates
the richness of the rotation involving asymmetric nuclear shape.

Moreover, the aplanar rotation in bands B4 and B4a is confirmed by
$K$-plots, and the azimuthal plots. Therefore, the planar and
aplanar rotations coexist in triaxial deformed nucleus $^{195}$Tl.

It is found that the asymmetric degree of the configuration $\pi
h_{9/2}\otimes \nu i^{-2}_{13/2}$ and the neutron holes in the
middle of $i_{13/2}$ shell favor the planar angular momentum
geometry within the $i$-$l$ plane instead of the traditional chiral
geometry. The present results suggest that the asymmetric degree of
the configuration in the nuclei of interest should be cautious in
the future study of nuclear chirality. Definitely, further
experimental efforts on the lifetime measurement of the states and
the comparison of transition probabilities with the calculation to
fully justify the rotational nature in $^{195}$Tl is helpful.

\section*{Acknowledgements}

The authors thank Professor G. Mukherjee for providing the numerical
data of experimental $B(M1)/B(E2)$ results and for reading of the
manuscript, and thank Professor J. Meng for helpful discussions.
This work is supported by the National Natural Science Foundation of
China (NSFC) under Grants No. 11775026 and 11875027, and the
Deutsche Forschungsgemeinschaft (DFG) and NSFC through funds
provided to the Sino-German CRC110 ``Symmetries and the Emergence of
Structure in QCD'' (DFG Grant No. TRR110 and NSFC Grant No.
11621131001).


\end{document}